\documentclass[journal, hidelinks]{IEEEtran}

\usepackage{amsmath,amssymb,amsfonts,mathtools}
\usepackage{algorithmic}
\usepackage{graphicx}
\usepackage{textcomp}
\usepackage{xcolor}
\usepackage{orcidlink}
\usepackage{cite}
\usepackage{enumitem}
\usepackage[toc, acronym]{glossaries}
\glsdisablehyper
\usepackage{comment}
\usepackage{balance}
\usepackage[final]{microtype}
\usepackage{scalerel}
\usepackage{siunitx}
\usepackage{balance}
\usepackage{pgfplots}
\pgfplotsset{compat=newest}
\usepgfplotslibrary{groupplots}
\usepgfplotslibrary{polar}
\usepgfplotslibrary{smithchart}
\usepgfplotslibrary{statistics}
\usepgfplotslibrary{dateplot}
\usepgfplotslibrary{ternary}
\usetikzlibrary{arrows.meta}
\usetikzlibrary{backgrounds}
\usepgfplotslibrary{patchplots}
\usepgfplotslibrary{fillbetween}

\let\originalleft\left
\let\originalright\right
\renewcommand{\left}{\mathopen{}\mathclose\bgroup\originalleft}
\renewcommand{\right}{\aftergroup\egroup\originalright}

\newcommand{\bsf}[1]{\ensuremath{\mathbf{#1}}}
\newcommand{\id}[1]{\ensuremath{\mathbf{I}_{#1}}}

\newcommand{\trans}[1]{\ensuremath{#1^{\mathrm{T}}}}

\newcommand{\Z}[1]{\ensuremath{\mathbf{Z}_{\mathrm{#1}}}}
\newcommand{\z}[1]{\ensuremath{\mathbf{z}_{\mathrm{#1}}}}
\newcommand{\ve}[1]{\ensuremath{\mathbf{v}_{\mathrm{#1}}}}
\newcommand{\iy}[1]{\ensuremath{\mathbf{i}_{\mathrm{#1}}}}
\newcommand{\ie}{\textit{i.e. }}
\newcommand{\e}[1]{\mathrm{e}^{\,#1}}
\newcommand{\im}{\ensuremath{\mathrm{j}}}

\newcommand{\frob}[1]{\ensuremath{\left\|#1\right\|_{\mathrm{F}}}}
\newcommand{\norm}[2]{\ensuremath{\|#1\|_{#2}}}

\usepackage{fancyhdr}
\newcommand{\changefont}{
	\color{blue}\fontsize{9}{9}\selectfont
}

\newcommand{\doi}{10.1109/LCOMM.2024.3416044}
\newcommand{\publication}{IEEE Communications Letters}
\let\oldmaketitle\maketitle
\renewcommand{\maketitle}{%
	\oldmaketitle
	\thispagestyle{fancy} 
}

\begin{document}
\newacronym{simo}{SIMO}{single-input multiple-output}
\newacronym{miso}{MISO}{multiple-input single-output}
\newacronym{mimo}{MIMO}{multiple-input multiple-output}
\newacronym{mmimo}{mMIMO}{massive MIMO}
\newacronym{snr}{SNR}{signal-to-noise ratio}
\newacronym{ula}{ULA}{uniform linear array}
\newacronym{ehf}{EHF}{extremely high frequency}
\newacronym{thf}{THF}{tremendously high frequency}
\newacronym{rhs}{RHS}{right-hand side}
\newacronym{lhs}{LHS}{left-hand side}

\title{Asymptotic Analysis of Near-Field Coupling in Massive MISO and Massive SIMO Systems}
\author{Aniol
	Martí\,\orcidlink{0000-0002-5600-8541},~\IEEEmembership{Graduate~Student~Member,~IEEE},
	Jaume Riba\,\orcidlink{0000-0002-5515-8169},~\IEEEmembership{Senior~Member,~IEEE},
	Meritxell~Lamarca\,\orcidlink{0000-0002-8067-6435},~\IEEEmembership{Member,~IEEE},
	and Xavier Gr\`acia\,\orcidlink{0000-0003-1006-4086}
	\thanks{This work was funded by projects MAYTE (PID2022-136512OB-C21) and GeMMFiT (PID2021-125515NB-C21) by MCIU/AEI/10.13039/501100011033 and ERDF ``A way of making Europe'', grants 2021 SGR 01033 and 2021 SGR 00603 by Departament de Recerca i Universitats de la Generalitat de Catalunya and grant 2023 FI ``Joan Or\'o'' 00050 by Departament de Recerca i Universitats de la Generalitat de Catalunya and the\nobreak\ ESF+.}%
	\thanks{A. Mart\'i. J. Riba and M. Lamarca are with the Departament de Teoria del Senyal i Comunicacions, Universitat Polit\`ecnica de Catalunya (UPC), 08034 Barcelona, Spain (e-mail: aniol.marti@upc.edu, jaume.riba@upc.edu, meritxell.lamarca@upc.edu).}%
	\thanks{X. Gr\`acia is with the Departament de Matem\`atiques, Universitat Polit\`ecnica de Catalunya (UPC), 08034 Barcelona, Spain (e-mail: xavier.gracia@upc.edu).}}

\maketitle

\begin{abstract}
This paper studies the receiver to transmitter antenna coupling in near-field communications with massive arrays.
Although most works in the literature consider that it is negligible and approximate it by zero, there is no rigorous analysis on its relevance for practical systems.
In this work, we leverage multiport communication theory to obtain conditions for the aforementioned approximation to be valid in \acrshort{miso} and \acrshort{simo} systems.
These conditions are then particularized for arrays with fixed inter-element spacing and arrays with fixed size.
\end{abstract}


\section{Introduction}
\IEEEPARstart{W}{ireless} communication systems up to 4G operate at frequency bands below \qty{5}{\GHz} with \acrshort{mimo} antenna configurations not larger than 8\texttimes8~\cite{3gpp_ts_2023}.
Under these specifications, the Fraunhofer distance~\cite[Sec.~2.2.4]{balanis_antenna_2016} is smaller than one meter, which means that communication always takes place in the far field.

However, the increasing demand of higher data rates together with the enormous growth of connected devices have pushed standards to adopt larger antenna configurations and frequencies in the \acrfull{ehf} and \acrfull{thf} bands.
For instance, in advanced releases of 5G, large arrays and carrier frequencies in the mmWave band are already employed, rising the far field distance to thousands of meters.
Furthermore, the attenuation at such frequencies makes it impossible to communicate in the far field, so near-field models must be taken into account.

In this paper, we focus on the coupling caused by the receiver to the transmitter when a massive array is employed in one of the communication system ends.
We analyze both the \acrshort{miso} and \acrshort{simo} scenarios, as they are important in the downlink and uplink, respectively.
In order to do so, we take advantage of \textit{multiport communication theory}~\cite{ivrlac_toward_2010, ivrlac_multiport_2014}, a framework that involves a circuit-theoretic approach where inputs and outputs of the \acrshort{mimo} communication system are associated with ports of a multiport black box, described by impedance or scattering matrices~\cite{tadele_channel_2023}.

Multiport models have shown their capability of incorporating non-ideal phenomena such as intra-array coupling in a more direct manner than electromagnetic information theory tools~\cite{chafii_twelve_2023, franceschetti_wave_2017}, which require a deep knowledge of both information and fields theory.
For instance, there are several works that rely on multiport theory to analyze different features of \acrshort{mimo} systems such as the array gain~\cite{ivrlac_receive_2009, laas_limits_2020}, the diversity gain~\cite{ivrlac_diversity_2011}, the multistreaming capability of compact arrays~\cite{ivrlac_multistreaming_2011}, the uplink/downlink reciprocity~\cite{laas_reciprocity_2020} and, more recently, the impact of mutual coupling in channel estimation~\cite{tadele_channel_2023} or to derive models for holographic \acrshort{mimo}~\cite{damico_holographic_2024}.
Nevertheless, most of these works are focused on far-field communication so unilateral coupling networks are considered~\cite[Sec.~11.2]{pozar_microwave_2012}.
A notable exception is~\cite{laas_limits_2020}, where the authors study the array gain of a \acrfull{mmimo} system operating in the near field, and assume that unilateral networks are no longer physically consistent in such scenario.
On the contrary, in this paper we focus on analyzing the conditions under which coupling from receiver antennas to transmitter antennas can be neglected in the near field.

In the following section, we review multiport communication theory and we derive a general noiseless input-output relation for the system.
Next, the system model considered in the study of inter-array coupling is presented in Section~\ref{sec:system_model}.
The rigorous analysis and its numerical evaluation for a finite number of antennas are presented in Sections~\ref{sec:study_coupling} and~\ref{sec:numerical_results}, respectively.
Finally, in Section~\ref{sec:conclusions}, we present our conclusions.

\section{Multiport Communication}
\label{sec:multiport}
\begin{figure*}
	\centering
	\includegraphics[width=\textwidth]{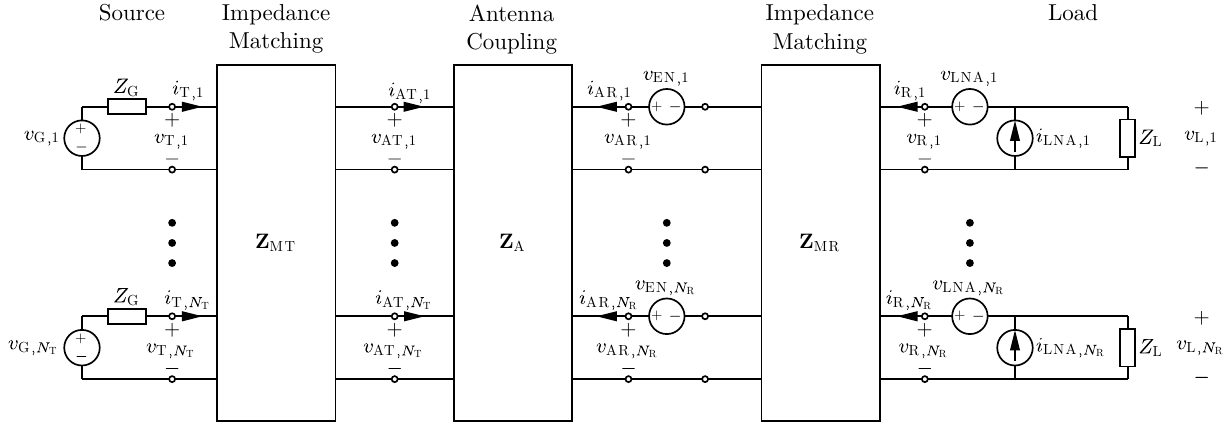}
	\caption{Multiport communication model of a $N_{\mathrm{T}} \times N_{\mathrm{R}}$ \acrshort{mimo} system. In the figure, $Z_{\mathrm{G}}$ and $Z_{\mathrm{L}}$ represent the generator and load impedances, respectively, and $v_{\mathrm{G},i}$ represents the generator source. Besides, $v_{\mathrm{EN},i}$, $v_{\mathrm{LNA},i}$ and $i_{\mathrm{LNA},i}$ stand for the noise sources (see \cite{ivrlac_toward_2010} for details).}
	\label{fig:multiport_mimo}
\end{figure*}

Consider a narrowband communication system with $N_{\mathrm{T}}$ antennas at the transmitter and $N_{\mathrm{R}}$ at the receiver.
The discrete baseband representation of the received signal is
\begin{equation}
	\bsf{y} = \bsf{H}\bsf{x} +\bsf{z},\quad \bsf{y},\bsf{z}\in\mathbb{C}^{N_{\mathrm{R}}},\, \bsf{x}\in\mathbb{C}^{N_{\mathrm{T}}},\, \bsf{H}\in\mathbb{C}^{N_{\mathrm{R}} \times N_{\mathrm{T}}},\label{eq:mimo_model}
\end{equation}
where $\bsf{z}$ is additive Gaussian noise and $\bsf{H}$ is the \acrshort{mimo} channel matrix that encompasses all the relevant physical context.
The multiport model for this system is shown in Fig.~\ref{fig:multiport_mimo}.
It consists of four basic parts: \textit{signal generation}, \textit{impedance matching}, \textit{antenna mutual coupling} and \textit{noise}. Since our goal is to analyze the coupling produced by the receiver on the transmitter in a \acrshort{mmimo} environment, we only focus on the second and third parts.
Further details on the multiport model and its relationship with~\eqref{eq:mimo_model} can be found in~\cite{ivrlac_toward_2010, ivrlac_multiport_2014, damico_holographic_2024}.

\subsection{Impedance matching}
\label{sec:impedance_matching}
Impedance matching networks are employed to maximize the power transfer to the antennas at the transmitter (\ie power matching) or to maximize the \acrfull{snr} at the receiver (\ie noise matching).
The transmitter matching network is described by its Z-parameters:
\begin{equation}
	\Z{MT} = \begin{pmatrix}
		\Z{MT,11} & \Z{MT,12}\\
		\Z{MT,21} & \Z{MT,22}
	\end{pmatrix} \in \mathbb{C}^{2N_{\mathrm{T}} \times 2N_{\mathrm{T}}}.
\end{equation}
The Z-parameters of the receiver-side matching network, $\Z{MR} \in \mathbb{C}^{2N_{\mathrm{R}} \times 2N_{\mathrm{R}}}$, are defined in a similar way.
Both networks are assumed to be \textit{lossless}, \textit{reciprocal} and \textit{noiseless}~\cite{ivrlac_toward_2010}.

When the number of antennas is large, designing optimal matching networks is quite demanding~\cite{ivrlac_toward_2010, ivrlac_multiport_2014, damico_holographic_2024}.
For this reason, in \acrshort{mmimo} systems, they are usually omitted~\cite{laas_limits_2020, laas_reciprocity_2020},
which translates into a lower \acrshort{snr} with no impact on coupling.

\subsection{Antenna mutual coupling}
\label{sec:antenna_coupling}
The antenna multiport Z-matrix allows for a compact description of mutual coupling and can be partitioned as:
\begin{equation}
	\Z{A} =
	\begin{pmatrix}
		\Z{AT} & \Z{ATR}\\
		\Z{ART} & \Z{AR}
	\end{pmatrix} \in \mathbb{C}^{(N_{\mathrm{T}}+N_{\mathrm{R}})\times(N_{\mathrm{T}}+N_{\mathrm{R}})},
\end{equation}
where $\Z{AT}$ and $\Z{AR}$ model the mutual coupling between antennas within the transmit or receive arrays (\ie intra-array coupling), respectively, and $\Z{ATR}$ and $\Z{ART}$ model the mutual coupling between the transmit and receive sides (\ie inter-array coupling).
Note that $\Z{A} = \trans{\Z{A}}$ due to antenna reciprocity~\cite[Sec.~3.8]{balanis_antenna_2016}.

When the transmitter and the receiver are sufficiently separated (\ie in the far field), it is usually assumed that the network is unilateral, so $\Z{ATR}=\mathbf{0}$.
In multiport communication theory, this is known as the \textit{unilateral approximation}~\cite{ivrlac_toward_2010} and implies that the currents at the receiver do not affect the transmitter.
Precisely, analyzing the validity of the unilateral approximation in the near field is the main goal of this paper.

\subsection{End-to-end (noiseless) system}
From Fig.~\ref{fig:multiport_mimo}, we define vectors $\ve{T} = \trans{(v_{\mathrm{T},1}, \dots, v_{\mathrm{T},N_{\mathrm{T}}})}$ and $\iy{T} = \trans{(i_{\mathrm{T},1}, \dots, i_{\mathrm{T},N_{\mathrm{T}}})}$, and equivalently for all currents and voltages in the figure.
If we now consider a noiseless scenario (\ie $\ve{EN}=\ve{LNA}=\iy{LNA}=\mathbf{0}$, $\ve{R} = \ve{L}$), the three multiports from Fig.~\ref{fig:multiport_mimo} can be combined into one:
\begin{equation}
	\begin{pmatrix}
		\ve{T}\\
		\ve{R}
	\end{pmatrix} =
	\begin{pmatrix}
		\Z{T}  & \Z{TR}\\
		\Z{RT} & \Z{R}
	\end{pmatrix}
	\begin{pmatrix}
		\iy{T}\\
		\iy{R}
	\end{pmatrix},
	\label{eq:multiport_input_output}
\end{equation}
where each block is obtained from circuit analysis as:
\begin{align}
	&\hspace*{-0.8em}\Z{T} \!= \Z{MT,11}\!-\!\Z{MT,12}\left(\mathbf{A}_{\mathrm{T}}\!-\!\Z{ATR}\mathbf{A}_{\mathrm{R}}^{-1}\Z{ART}\right)^{-1}\Z{MT,21},\nonumber\\
	&\hspace*{-0.8em}\Z{R} \!=\! \Z{MR,11}\!-\!\Z{MR,12}\left(\mathbf{A}_{\mathrm{R}}\!-\!\Z{ART}\mathbf{A}_{\mathrm{T}}^{-1}\Z{ATR}\right)^{-1}\Z{MR,21},\!\label{eq:z_r}\\
	&\hspace*{-0.8em}\Z{TR} \!=\! \Z{MT,12}\left(\mathbf{A}_{\mathrm{T}}\!-\!\Z{ATR}\mathbf{A}_{\mathrm{R}}^{-1}\Z{ART}\right)^{-1}\Z{ATR}\mathbf{A}_{\mathrm{R}}^{-1}\Z{MR,21},\nonumber
\end{align}
with the following auxiliary variables:
\begin{equation}
	\mathbf{A}_{\mathrm{R}} = \Z{AR}+\Z{MR,22},\quad \mathbf{A}_{\mathrm{T}} = \Z{AT}+\Z{MT,22}.
\end{equation}
This network is also reciprocal, as it is a cascade connection of three reciprocal networks~\cite{anderson_reciprocity_1965}, hence:
\begin{equation}
	\Z{T} = \trans{\Z{T}},\quad \Z{R} = \trans{\Z{R}},\quad \Z{RT} = \trans{\Z{TR}}.
\end{equation}

From Eq.~\eqref{eq:multiport_input_output} and Ohm's law, we can obtain the matrix of the (noiseless) input-output relationship $\ve{L}=\mathbf{D}\ve{G}$:
\begin{equation}
	\label{eq:matrix_D}
	\begin{multlined}
		\mathbf{D}=Z_{\mathrm{L}}\left(Z_{\mathrm{L}}\id{}+\Z{R}\right)^{-1}\Z{RT}\\
		\times(Z_{\mathrm{G}}\id{}+\Z{T}-\Z{TR}(Z_{\mathrm{L}}\id{}+\Z{R})^{-1}\Z{RT})^{-1}.
	\end{multlined}
\end{equation}
Applying the matrix identity derived in the appendix, $\mathbf{D}$ can also be expressed as:
\begin{equation}
	\label{eq:matrix_D_2}
	\begin{multlined}
		\mathbf{D}=Z_{\mathrm{L}}\left(Z_{\mathrm{L}}\id{}+\Z{R}-\Z{RT}(Z_{\mathrm{G}}\id{}+\Z{T})^{-1}\Z{TR}\right)^{-1}\\
		\times\Z{RT}(Z_{\mathrm{G}}\id{}+\Z{T})^{-1}.
	\end{multlined}
\end{equation}

Recalling the unilateral approximation described in~\ref{sec:antenna_coupling} and applying it into~\eqref{eq:z_r} results in $\Z{TR}=0$, which implies that~\eqref{eq:multiport_input_output} is also a unilateral network.
This assumption simplifies the analysis greatly and leads to the following input-output matrix:
\begin{equation}
	\label{eq:d_unilateral}
	\mathbf{D}_{\mathrm{UA}}=Z_{\mathrm{L}}\left(Z_{\mathrm{L}}\id{}+\Z{R}\right)^{-1}\Z{RT}(Z_{\mathrm{G}}\id{}+\Z{T})^{-1}.
\end{equation}
However, this is a coarse approximation.
Actually, for~\eqref{eq:d_unilateral} to be valid --and therefore for the system to behave as if the network was unilateral-- we do not need $\Z{TR}=\mathbf{0}$ but only inter-array coupling to be much smaller than intra-array coupling.
Denoting the Frobenius norm~\cite[Sec.~5.6]{horn_matrix_2017} with~$\frob{\boldsymbol{\cdot}}$, the condition can be expressed as:
\begin{equation}
	\label{eq:condition_1}
	\frob{\Z{TR}(Z_{\mathrm{L}}\id{}+\Z{R})^{-1}\Z{RT}} \ll \frob{Z_{\mathrm{G}}\id{}+\Z{T}},
\end{equation}
or equivalently as
\begin{equation}
	\frob{\Z{RT}(Z_{\mathrm{G}}\id{}+\Z{T})^{-1}\Z{TR}} \ll \frob{Z_{\mathrm{L}}\id{}+\Z{R}},
\end{equation}
since $\mathbf{D} \approx \mathbf{D}_{\mathrm{UA}}$ in both cases.

Finally, we shall remark that when the unilateral approximation does not hold, matching networks design becomes unfeasible due to the dependence between $\Z{MT}$ and $\Z{MR}$ observed in~\eqref{eq:z_r}.
This fact further justifies their omission in \acrshort{mmimo} systems.
For this reason, from now on we will consider that there are no matching networks, without any detriment on the validity of the asymptotic analysis.

\section{System Model}
\label{sec:system_model}
We consider two complementary wireless communication systems (\ie downlink and uplink):
\begin{enumerate}[label=\Alph*)]
	\item Massive \acrshort{miso} with $N_{\mathrm{T}}=N$ and $N_{\mathrm{R}}=1$.
	\item Massive \acrshort{simo} with $N_{\mathrm{T}}=1$ and $N_{\mathrm{R}}=N$.
\end{enumerate}
The array is assumed to be a \acrfull{ula} with aperture size $D$ and
radiating elements are Hertzian dipoles of length $l$ with a separation $d$ between them.
We assume that loading coils are inserted between every antenna and its feedline to compensate for the capacitive reactance of the dipoles~\cite[Ch.~2]{pozar_microwave_2012}.
With this procedure, the intra-array coupling impedance matrix is given by:
\begin{equation}
	\begin{cases}
		[\Z{\boldsymbol{\cdot}}]_{m,n} = R_{\mathrm{r}},\ m=n,\ 1\leq m,n \leq N,\\
		[\Z{\boldsymbol{\cdot}}]_{m,n} = R_{\mathrm{r}}\psi\left(kd|m\!-\!n|\right),\ m\neq n,
	\end{cases}
	\label{eq:intraarray_coupling}
\end{equation}
where $\boldsymbol{\cdot}$ denotes either $\mathrm{T}$ or $\mathrm{R}$ as appropriate, $R_{\mathrm{r}}=\frac{2}{3}\pi \eta (\frac{l}{\lambda})^2$ is the radiation resistance, with $\eta$ the impedance of free space, $k=2\pi/\lambda$ is the wavenumber and $\psi$ is a function proportional to the electric field of a Hertzian dipole~\cite[Sec.~4.2]{balanis_antenna_2016},~\cite{yordanov_arrays_2009},~\cite{bjornson_primer_2021}:
\begin{equation}
	\label{eq:psi}
	\psi(x) = \frac{3}{2}\,\im\,\e{-\im x}\left(\frac{1}{x}-\frac{\im}{x^2}-\frac{1}{x^3}\right).
\end{equation}

Regarding inter-array coupling, if it is assumed that communication takes place in the \textit{radiative} near field, transmitting antennas can be considered as point sources.
Taking into account that the expression for inter-array coupling degenerates into a vector in both \acrshort{miso} and \acrshort{simo}, we obtain:
\begin{equation}
	[\z{TR}]_n = \complexnum{j} R_{\mathrm{r}} \frac{\e{-\complexnum{j}kr_n}}{kr_n},
\end{equation}
where $r_n$ is the distance between the single-antenna device and the $n$-th element in the array.
Observe that this model is consistent with~\eqref{eq:psi} but neglecting the last two terms~\cite{bjornson_primer_2021}.

\section{Study of Coupling Effects}
\label{sec:study_coupling}
We now proceed to analyze the inter-array coupling in the \acrshort{miso} and \acrshort{simo} scenarios described in the previous section.
For ease of reading, we summarize the system parameters:
\begin{itemize}
	\item Number of antennas: $N$.
	\item Array size: $D$.
	\item Inter-element spacing: $d$.
	\item Dipole length: $l$.
	\item Wavelength: $\lambda$.
	\item Distance between transmitter and receiver: $r$.
	\item Incidence angle: $\theta$.
	\item Source and load impedances: $Z_{\mathrm{G}}$, $Z_{\mathrm{L}}$.
\end{itemize}
Throughout this section we will use the following abbreviations:
\begin{equation}
	\begin{aligned}
		\Z{GT} = Z_{\mathrm{G}}\id{}+\Z{T}&,\quad \Z{RL} = Z_{\mathrm{L}}\id{}+\Z{R},\\
		Z_{\mathrm{GT}} = Z_{\mathrm{G}} + R_{\mathrm{r}}&,\quad Z_{\mathrm{RL}} =  Z_{\mathrm{L}} + R_{\mathrm{r}}.
	\end{aligned}	
\end{equation}

\subsection{Massive MISO}
\label{sec:miso}
In a \acrshort{miso} system, Eq.~\eqref{eq:matrix_D} becomes:
\begin{equation}
	\label{eq:d_miso}
	\trans{\mathbf{d}} = \frac{Z_{\mathrm{L}}}{Z_{\mathrm{RL}}}\trans{\mathbf{z}_{\mathrm{TR}}}
	\left(\Z{GT}-\frac{\mathbf{z}_{\mathrm{TR}}\trans{\mathbf{z}_{\mathrm{TR}}}}{Z_{\mathrm{RL}}}\right)^{-1}.
\end{equation}
The unilateral approximation will be valid when the receiver has no effect on the transmitter.
That is, when condition~\eqref{eq:condition_1} is fulfilled:
\begin{equation}
	\frob{\frac{\mathbf{z}_{\mathrm{TR}}\trans{\mathbf{z}_{\mathrm{TR}}}}{Z_{\mathrm{RL}}}} \ll \frob{\Z{GT}}.
\end{equation}
Since $\z{TR}\trans{\z{TR}}$ is rank-one, it is equivalent to:
\begin{equation}
	\label{eq:miso_condition}
	\frac{\norm{\z{TR}}{2}^2}{|Z_{\mathrm{RL}}|} \ll \frob{\Z{GT}}.
\end{equation}

Assuming the intra-array coupling model in~\eqref{eq:intraarray_coupling}, we can further manipulate the \acrfull{rhs} in~\eqref{eq:miso_condition}.
In particular, we lower bound $\frob{\Z{GT}}$ as follows:
\begin{multline}
	\frob{\Z{GT}} = \sqrt{\sum_{m=1}^{N}\sum_{n=1}^{N}|[\Z{GT}]_{m,n}|^2} = \sqrt{\sum_{m=1}^{N}\norm{[\Z{GT}]_{m, *}}{2}^2}\\
	\geq \sqrt{N}\cdot\min_{m}\norm{[\Z{GT}]_{m, *}}{2} = \sqrt{N}\cdot\norm{[\Z{GT}]_{1, *}}{2}.	\label{eq:frob_bound}
\end{multline}
In the last step, we have used that $|\psi(x)|^2$ is monotonically decreasing with $x>0$.
It is also worth to remark that $\sqrt{N}\,\norm{[\Z{GT}]_{1, *}}{2} \to \frob{\Z{GT}}$ when $N \to \infty$, because $\Z{GT}$ is Toeplitz. Therefore, the bound in~\eqref{eq:frob_bound} is asymptotically tight.

We now proceed to compute the Euclidean norm of the first row:
\begin{equation}
	\begin{aligned}
		&\norm{[\Z{GT}]_{1, *}}{2}^2 = |Z_{\mathrm{GT}}|^2+\sum_{n=1}^{N-1}R_{\mathrm{r}}^2|\psi(kdn)|^2\\
		&= |Z_{\mathrm{GT}}|^2+\frac{9}{4}R_{\mathrm{r}}^2\sum_{n=1}^{N-1}\frac{1}{(kdn)^2}-\frac{1}{(kdn)^4}+\frac{1}{(kdn)^6}\\
		&= |Z_{\mathrm{GT}}|^2+\frac{9}{4}R_{\mathrm{r}}^2\left(\frac{H_{N-1}^{(2)}}{(kd)^2}-\frac{H_{N-1}^{(4)}}{(kd)^4}+\frac{H_{N-1}^{(6)}}{(kd)^6}\right),
	\end{aligned}
\end{equation}
where $H_{k}^{(q)}$ is the generalized harmonic number of order $q$~\cite[Sec.~1.2.7]{knuth_art_1997}.
Multiplying by $N$ we obtain the bound in~\eqref{eq:frob_bound}:
\begin{equation}
	\label{eq:zt_miso}
	\begin{multlined}
		\frob{\Z{GT}}^2 \geq N|Z_{\mathrm{GT}}|^2\\
		+\frac{9}{4}NR_{\mathrm{r}}^2\left(\frac{H_{N-1}^{(2)}}{(kd)^2}-\frac{H_{N-1}^{(4)}}{(kd)^4}+\frac{H_{N-1}^{(6)}}{(kd)^6}\right).
	\end{multlined}
\end{equation}
Substituting the square root of~\eqref{eq:zt_miso} in~\eqref{eq:miso_condition} results in a sufficient condition for the unilateral approximation to hold in \acrshort{miso}.

Now we analyze this condition in two limiting cases: fixed inter-element spacing and fixed array size.
\subsubsection{Fixed inter-element spacing}
If $d$ is fixed, then $\frob{\Z{GT}}$ diverges as $O(\sqrt{N})$.
Furthermore, it can be assumed that the receiver is located at a distance $r$ meters and incidence angle $\theta=\frac{\pi}{2}$ to the first antenna in the transmitting array, as the asymptotic behavior does not depend on the location of the receiving antenna.
Under these assumptions, the \acrfull{lhs} in~\eqref{eq:miso_condition} can be written as:
\begin{equation}
	\norm{\z{TR}}{2}^2 = \frac{R_{\mathrm{r}}^2}{k^2} \sum_{n=0}^{N-1} \frac{1}{r_n^2} = \frac{R_{\mathrm{r}}^2}{k^2} \sum_{n=0}^{N-1} \frac{1}{r^2+(nd)^2}.
	\label{eq:miso_series}
\end{equation}
The limit when $N\to\infty$ of the sum above is convergent and can be computed using the Poisson summation formula~\cite[Sec.~7.3]{beerends_fourier_2003}:
\begin{equation}
	\label{eq:limit_miso}
	\hspace*{-0.075em}\lim_{N\to+\infty}\frac{\norm{\z{TR}}{2}^2}{|Z_{\mathrm{RL}}|} = \frac{R_{\mathrm{r}}^2}{|Z_{\mathrm{RL}}|}\left(\frac{d+r\pi\coth\left(\pi\frac{r}{d}\right)}{2d(rk)^2}\right) < +\infty,
\end{equation}
which means that \eqref{eq:miso_condition} is true asymptotically.
Therefore, in a massive \acrshort{miso} environment, the unilateral approximation is asymptotically valid even in the (radiative) near field when the inter-element spacing is kept constant.

\subsubsection{Fixed array size}
In order to keep the array size constant, we must let the inter-element distance decrease with the number of antennas: $d=\frac{D}{N-1}$.
Substituting it in \eqref{eq:zt_miso}, we observe that $\frob{\Z{GT}}$ grows as $O(N^{3.5})$, compared with $O(\sqrt{N})$ from the previous scenario.
Note that this is consistent with the fact that placing antennas closer together increases mutual coupling.

Besides, substituting $d$ in~\eqref{eq:miso_series}, we obtain that each term of the series is given by:
\begin{equation}
	a_n = \frac{R_{\mathrm{r}}^2/k^2}{r^2+(\frac{nD}{N-1})^2} \geq \frac{R_{\mathrm{r}}^2/k^2}{r^2+(\frac{nD}{n})^2} = \frac{R_{\mathrm{r}}^2/k^2}{r^2+D^2},
\end{equation}
and applying the vanishing condition and the comparison test~\cite[Ch.~23]{spivak_1994}, it follows that $\norm{\z{TR}}{2}^2$ diverges as $O(N)$.
As the \acrshort{rhs} in~\eqref{eq:miso_condition} grows much faster than its \acrshort{lhs}, we conclude that the unilateral approximation is also asymptotically valid when the array size is fixed.

\subsection{Massive SIMO}
\label{sec:simo}
The (noiseless) input-output system vector in \acrshort{simo} is:
\begin{equation}
	\label{eq:d_simo_2}
	\mathbf{d} = \frac{Z_{\mathrm{L}}}{Z_{\mathrm{GT}}} \left(\Z{RL}-\frac{\mathbf{z}_{\mathrm{RT}}\trans{\mathbf{z}_{\mathrm{RT}}}}{Z_{\mathrm{GT}}}\right)^{-1}\mathbf{z}_{\mathrm{RT}},
\end{equation}
which has been derived from~\eqref{eq:matrix_D_2}.
Hence, the unilateral approximation condition becomes:
\begin{equation}
	\frac{\norm{\z{RT}}{2}^2}{|Z_{\mathrm{GT}}|} \ll \frob{\Z{RL}},\label{eq:simo_condition}
\end{equation}
which is equivalent to~\eqref{eq:miso_condition} derived for \acrshort{miso}.

If the system considered in Sec.~\ref{sec:miso} was now employed as a \acrshort{simo} system in the uplink, then the expressions derived for \acrshort{miso} (\ie \eqref{eq:miso_condition} to \eqref{eq:limit_miso}) would be valid if we replaced $\Z{GT}$ and $Z_{\mathrm{RL}}$ with $\Z{RL}$ and $Z_{\mathrm{GT}}$.
This allows to extend the downlink results to the uplink, which means that the unilateral approximation is also asymptotically valid for a \acrshort{simo} system operating in the radiative near field.

\section{Numerical Results}
\label{sec:numerical_results}
In this section, the theoretical results developed in Section~\ref{sec:study_coupling} are numerically illustrated in two scenarios with realistic parameters.
In the first one, we consider a \acrshort{ula} with fixed inter-element spacing $d=\lambda/2$ at a distance $r=\SI{55}{\m}$, which imposes a maximum aperture $D_{\mathrm{max}}=\SI{2}{\m}$ in order to operate in the Fresnel region.
In the second scenario, a \acrshort{ula} with constant aperture $D=\SI{1}{\m}$ and a separation between transmitter and receiver equal to the Fresnel distance $r=0.62\sqrt{D^3/\lambda}\approx\SI{20}{\m}$ are considered.
In both cases, the system operates at $\lambda=\SI{1}{\mm}$ and the dipole length is $l=\lambda/20$.
The source and load impedances $Z_{\mathrm{G}} = Z_{\mathrm{L}} = \complexnum{186 - 31.6i}\,\unit{\ohm}$ were measured in~\cite{lehmeyer_lna_2015} and have been recently used in~\cite{damico_holographic_2024}.
A summary of this setup is given in Table~\ref{tab:setup}.
For the sake of clarity, in this section we employ the notation for \acrshort{miso} systems, although results are also valid for \acrshort{simo} systems, as they have been shown to be equivalent in Sec.~\ref{sec:simo}.
\begin{table}[htb]
	\renewcommand{\arraystretch}{1.21}
	\caption{Summary of the parameters employed in the simulations}
	\label{tab:setup}
	\centering
	\begin{tabular}{c|cc}
		Parameter & Scenario 1                    & Scenario 2              \\ \hline\hline
		$N$       & $[10, 2000]$                  & $[10, 10^6]$            \\ \hline
		$D$       & $[\SI{4.5}{\mm}, \SI{2}{\m}]$ & $\SI{1}{\m}$            \\ \hline
		$d$       & $\lambda/2$                   & $\frac{D}{N-1}$         \\ \hline
		$l$       & $\lambda/20$                  & $\lambda/20$            \\ \hline
		$\lambda$ & $\SI{1}{\mm}$                 & $\SI{1}{\mm}$           \\ \hline
		$r$       & $\SI{55}{\m}$                 & $\SI{20}{\m}$           \\ \hline
		$\theta$  & $\pi/2\,\unit{\radian}$       & $\pi/2\,\unit{\radian}$ \\ \hline
		$Z_{\mathrm{G}}$, $Z_{\mathrm{L}}$ & $\complexnum{186 - 31.6i}\,\unit{\ohm}$ & $\complexnum{186 - 31.6i}\,\unit{\ohm}$
	\end{tabular}
\end{table}

\begin{figure}[htb]
	\centering
	\includegraphics[width=\columnwidth]{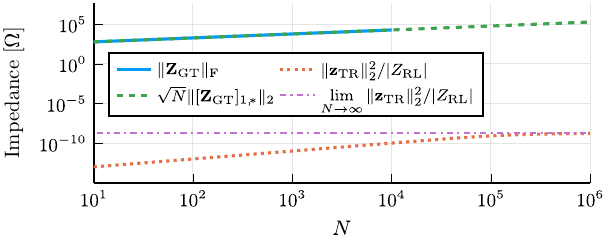}
	\caption{Fixed inter-element spacing $d=\lambda/2$ coupling condition as a function of $N$ with $r=\SI{55}{\m}$. $\frob{\Z{GT}}$ is displayed up to $N=10^4$ due to computational limitations.}
	\label{fig:fixed_spacing}
\end{figure}

In Fig.~\ref{fig:fixed_spacing} the results for the first scenario are shown.
We depict $\frob{\Z{GT}}$ and its lower bound from~\eqref{eq:frob_bound}, as well as~\eqref{eq:miso_series} and~\eqref{eq:limit_miso}.
It can be observed that, for this particular setup, the unilateral approximation holds for any practical number of antennas, as $\norm{\mathbf{z}_{\mathrm{TR}}}{2}^2/|Z_{\mathrm{RL}}|$ is always much smaller (at least 10 times) than $\frob{\Z{GT}}$.
It should be noted that the results for $N>2000$ are not valid for our model, as the system would be operating in the inductive near field, but they are drawn to illustrate the convergence of~\eqref{eq:miso_series} into~\eqref{eq:limit_miso}.

\begin{figure}[htb]
	\centering
	\includegraphics[width=\columnwidth]{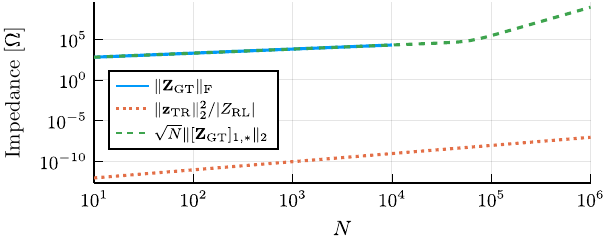}
	\caption{Fixed array size $D=\SI{1}{\m}$ coupling condition as a function of $N$ with $r\approx\SI{20}{\m}$ (\ie the Fresnel distance).  $\frob{\Z{GT}}$ is displayed up to $N=10^4$ due to computational limitations.}
	\label{fig:fixed_aperture}
\end{figure}

In Fig.~\ref{fig:fixed_aperture}, we plot the results for the second simulation scenario.
As expected, $\norm{\mathbf{z}_{\mathrm{TR}}}{2}^2/|Z_{\mathrm{RL}}|$ grows without bounds, but $\frob{\Z{GT}}$ grows much faster, thus ensuring the asymptotic validity of the unilateral approximation.
Furthermore, for all $N$ it is fulfilled $\norm{\mathbf{z}_{\mathrm{TR}}}{2}^2/|Z_{\mathrm{RL}}| \ll \frob{\Z{GT}}$.
Therefore, inter-array coupling can also be neglected in this scenario.

Finally, we shall remark that the lower bound for $\frob{\Z{GT}}$ derived in~\eqref{eq:frob_bound} is tight in both scenarios.

\section{Conclusions}
\label{sec:conclusions}
Exploiting multiport communication theory, we have derived the (noiseless) input-output equation of an arbitrary \acrshort{mmimo} system in terms of impedance matrices.
Next, we have established a condition for near-field coupling from the receiver to the transmitter to be negligible.
We have analyzed it under an asymptotic regime when a \acrshort{ula} is employed at the transmitter (\acrshort{miso}, downlink) or at the receiver (\acrshort{simo}, uplink).
We have shown that the same condition is valid both for \acrshort{miso} and \acrshort{simo} if the system considered is the same but operating in the downlink or the uplink, respectively.
In the analysis, we have focused on two limiting cases: fixed array size and fixed inter-element spacing.
In both cases, receiver to transmitter coupling can be neglected when communication takes place in the far field or the radiative near field.

\appendix
Consider nonsingular matrices $\mathbf{T} \in \mathbb{C}^{N\times N}$, $\mathbf{R} \in \mathbb{C}^{M\times M}$ and matrices $\mathbf{Z} \in \mathbb{C}^{M\times N}$, $\mathbf{Y} \in \mathbb{C}^{N\times M}$.
We shall prove the following identity:
\begin{equation}
	\label{eq:appendix_equality}
	\mathbf{R}^{-1}\mathbf{Z}(\mathbf{T}-\mathbf{Y}\mathbf{R}^{-1}\mathbf{Z})^{-1} = (\mathbf{R}-\mathbf{Z}\mathbf{T}^{-1}\mathbf{Y})^{-1}\mathbf{Z}\mathbf{T}^{-1}.
\end{equation}
Applying Woodbury's identity~\cite[Sec.~0.7]{horn_matrix_2017},
\begin{equation}
	\begin{multlined}
		\mathbf{R}^{-1}\mathbf{Z}(\mathbf{T}-\mathbf{Y}\mathbf{R}^{-1}\mathbf{Z})^{-1}\\
		= \mathbf{R}^{-1}\mathbf{Z}(\mathbf{T}^{-1}+\mathbf{T}^{-1}\mathbf{Y}(\mathbf{R}-\mathbf{Z}\mathbf{T}^{-1}\mathbf{Y})^{-1}\mathbf{Z}\mathbf{T}^{-1}).
	\end{multlined}
\end{equation}
Right-factoring out $(\mathbf{R}-\mathbf{Z}\mathbf{T}^{-1}\mathbf{Y})^{-1}\mathbf{Z}\mathbf{T}^{-1}$,
\begin{equation}
	\begin{aligned}
		\mathbf{R}^{-1}(\mathbf{R}-\mathbf{Z}\mathbf{T}^{-1}\mathbf{Y}+&\mathbf{Z}\mathbf{T}^{-1}\mathbf{Y})(\mathbf{R}-\mathbf{Z}\mathbf{T}^{-1}\mathbf{Y})^{-1}\mathbf{Z}\mathbf{T}^{-1}\\
		&= (\mathbf{R}-\mathbf{Z}\mathbf{T}^{-1}\mathbf{Y})^{-1}\mathbf{Z}\mathbf{T}^{-1},
	\end{aligned}
\end{equation}
which concludes the proof.\hfill\IEEEQED

\section*{Acknowledgment}
The authors would like to thank Profs.~Jordi Romeu, Juan M. Rius and Albert Aguasca from Universitat Politècnica de Catalunya for their helpful comments.

\bibliographystyle{IEEEtran}
\bibliography{IEEEabrv,references}
\end{document}